\documentclass[10pt,twocolumn]{article}

\usepackage[T1]{fontenc}
\usepackage[utf8]{inputenc}
\usepackage{graphicx}
\usepackage{amsmath,amssymb}
\usepackage{natbib}
\usepackage[margin=0.75in]{geometry}
\usepackage[hidelinks]{hyperref}
\usepackage{microtype}

\newcommand{\arcsec}{\ensuremath{^{\prime\prime}}}

\title{Phase Separation of HI in the Draco Nebula: CNM formation Triggered by Colliding WNM Flows}

\author{%
Yuhi \textsc{NARUSE}$^{1,*}$, Hiroaki \textsc{YAMAMOTO}$^{1}$, Yamato \textsc{MATSUZUKI}$^{1}$,\\
Rin \textsc{YAMADA}$^{2}$, Atsushi \textsc{NISHIMURA}$^{2}$, and Toshikazu \textsc{ONISHI}$^{3}$\\[0.7em]
\small $^{1}$Department of Physics, Graduate School of Science, Nagoya University,\\
\small Furo-cho, Chikusa-ku, Nagoya, Aichi 464-8602, Japan\\
\small $^{2}$Nobeyama Radio Observatory, National Astronomical Observatory of Japan, National Institutes of Natural Sciences,\\
\small Nobeyama, Minamimaki, Minamisaku, Nagano 384-1305, Japan\\
\small $^{3}$Department of Physics, Graduate School of Science, Osaka Metropolitan University,\\
\small 3-3-138 Sugimoto, Sumiyoshi-ku, Osaka 558-8585, Japan\\
\small $^{*}$E-mail: \texttt{y.naruse@a.phys.nagoya-u.ac.jp}%
}
\date{}

\begin{document}
\maketitle
\begin{abstract}

We investigate the thermal phase structures and kinematics of atomic hydrogen
(H\,\textsc{i}) gas in the Draco nebula to clarify how infalling
H\,\textsc{i} evolves toward cold atomic and molecular clouds.
Using the DHIGLS H\,\textsc{i} 21~cm data, we apply the Regularized Optimization
for Hyper-Spectral Analysis (ROHSA) algorithm to decompose the emission into
spatially coherent Gaussian components, and classify them as warm neutral
medium (WNM), lukewarm neutral medium (LNM), and cold neutral medium (CNM).
We identify two kinematically distinct WNM components: WNM1 near
$-3~\mathrm{km\,s^{-1}}$ and WNM2 near
$-30~\mathrm{km\,s^{-1}}$.
The LNM forms a broad velocity bridge between two WNMs, whereas the CNM is
concentrated within a narrower intermediate velocity range.
CO emission is also confined to this interval and is associated with
CNM-rich regions.
The H\,\textsc{i} mass fractions are $\sim41\%$, $\sim23\%$, $\sim17\%$,
and $\sim19\%$ for WNM1, WNM2, LNM, and CNM, respectively, with the LNM
and CNM together accounting for $\sim36\%$ of the total H\,\textsc{i} mass.
No comparably independent WNM component is recovered at intermediate
velocities, although a weak broad component may remain unresolved.
The observed phase structures are consistent with a dynamically compressed
multiphase interaction layer in which compression increases the gas density,
shortens the radiative cooling time, and drives the evolution from WNM through
thermally unstable gas toward the CNM.
The present interaction may both create new CNM, and further compress or grow
pre-existing dense structures.
Our results provide the missing link between previous evidence for dynamical
interaction and molecular cloud formation in Draco, and identify CNM formation
as a key step in transforming gas accreting toward the Galactic disk into cold
atomic and molecular clouds.

\end{abstract}

\medskip

\section{Introduction}

The formation of cold atomic and molecular gas from the diffuse interstellar medium is a fundamental process in the evolution of the interstellar medium and in the formation of molecular clouds.  Neutral atomic hydrogen (H\,\textsc{i}) is commonly described in terms of two thermally stable phases, the cold neutral medium (CNM) and the warm neutral medium (WNM), which can coexist over a limited range of thermal pressures \citep{Field1969,Wolfire2003}.  Between these two stable regimes, gas can occupy thermally unstable temperatures.  H\,\textsc{i} emission and absorption observations have shown that a substantial fraction of the WNM lies in this thermally unstable regime, demonstrating that the neutral atomic medium cannot be described solely by the two stable phases (e.g., \cite{Heiles2003}).  The conversion between the WNM, thermally unstable gas, and CNM is therefore an important element of the thermal evolution of the atomic interstellar medium.

The thermal state of H\,\textsc{i} gas can be strongly modified by dynamical compression. When WNM is compressed, its density increases, and its radiative cooling time decreases, allowing the H\,\textsc{i} gas to leave the stable warm phase, and enter the thermally unstable regime.  If radiative cooling proceeds sufficiently rapidly compared with the relevant dynamical timescale, condensation can subsequently produce CNM.  Analytical and numerical studies have shown that compression by shocks or converging H\,\textsc{i} flows can efficiently drive this transition \citep{Hennebelle1999,Koyama2000,Audit2005,Saury2014}. Numerical simulations of converging WNM flows further show that the resulting interaction layers are intrinsically multiphase, containing stable WNM, thermally unstable gas, and CNM simultaneously (e.g., \cite{Micic2013,Kobayashi2020,Kobayashi2022}).  The relative amounts of these phases depend on the properties of the inflowing gas, shock geometry, cooling efficiency, and turbulence generated within the interaction region.  Therefore, CNM formation in dynamically compressed H\,\textsc{i} gas can provide an important intermediate step between diffuse WNM and molecular cloud formation.

Observational identification of this phase transition is challenging because multiple H\,\textsc{i} components with different thermal and kinematic properties often overlap along the same line of sight.  H\,\textsc{i} 
absorption measurements provide direct constraints on cold atomic gas, and can also constrain warmer components, when combined with emission data, but such measurements are limited to sight lines toward suitable continuum sources.  Therefore, H\,\textsc{i} emission surveys in large area provide an important complementary means of studying the phase structure of the neutral medium.  Gaussian decomposition of H\,\textsc{i} emission has been widely used to statistically separate components with different characteristic line widths. \citet{Kalberla2018}, using a Gaussian decomposition of the all-sky HI4PI data, identified characteristic linewidth regimes associated with the CNM, thermally unstable lukewarm neutral medium (LNM), and WNM.  More recently, \citet{Matsuzuki2026} applied ROHSA (the Regularized Optimization for Hyper-Spectral Analysis) to GALFA-HI data of the molecular cloud forming H\,\textsc{i} clouds, HLCG 92--35, and separated the H\,\textsc{i} emission into WNM, LNM, and CNM components, revealing a large thermally unstable fraction and sub parsec CNM structures.  Spatially regularized methods such as ROHSA further allow spectrally overlapping components to be decomposed while preserving their spatial coherence across the data cube \citep{Marchal2019}.  Such approaches provide a powerful means of connecting the thermal and kinematic structures of H\,\textsc{i} over extended regions.

Intermediate velocity clouds (IVCs) offer particularly useful environments to investigate the dynamical conversion of diffuse H\,\textsc{i} gas.  IVCs are thought to be gas circulated between the Galactic disk and halo, and many are likely related to Galactic fountain processes that material expelled from the disk subsequently returns toward the Galactic plane (e.g., \cite{Wakker2001,Rohser2016}).  Interactions between infalling intermediate velocity gas and lower velocity Galactic gas can generate compression, shocks, and turbulence, creating conditions favorable for thermal instability and the formation of cold atomic gas.  Molecular hydrogen has been detected along a substantial number of IVC sight lines, indicating that cold and sufficiently dense substructures are present within at least some IVC complexes (e.g., \cite{Richter2003}).  \citet{Rohser2016} further identified a population of candidates of molecular intermediate velocity cloud at high Galactic latitude. IVCs provide a valuable laboratory for studying the connection between large scale gas dynamics, atomic phase conversion, and the development of molecular material.

The Draco nebula is one of the most prominent intermediate velocity H\,\textsc{i} structures situated at high Galactic latitude, and has been studied in H\,\textsc{i} since early observations by \citet{Goerigk1983}.  More recent observations have revealed that Draco contains a highly structured mixture of atomic, cold dense, and molecular gas. Herschel observations by \citet{MivilleDeschenes2017} revealed a fragmented network of dense dust structures, and were interpreted as evidence that the infalling cloud is interacting dynamically with diffuse Galactic gas.  They further suggested that Rayleigh--Taylor instabilities, and turbulent dissipation play important roles in shaping the interaction interface.  CO observations by \citet{BiegingKong2024} revealed highly clumpy molecular clouds with a complex velocity field, indicating that the molecular component is dynamically structured rather than quiescent.  \citet{Schneider2024} detected C\,\textsc{ii} emission at intermediate velocities, and found that the observed intensities are difficult to explain solely by photodissociation under the weak far ultraviolet radiation field in Draco.  They proposed an additional contribution from mechanical heating associated with shocks.  \citet{Schneider2025} further argued that Draco is undergoing a non-equilibrium H\,\textsc{i} to H$_{2}$ transition, and is likely at an early stage of molecular cloud evolution.

Despite this extensive evidence for dynamical interaction and ongoing molecular evolution, the detailed thermal phase structure of the atomic gas in Draco remains less well constrained. In particular, the kinematic relationship between the WNM, LNM, and CNM has not yet been established over the cloud as a whole.  It remains unclear whether distinct warm atomic components are the components related to the interaction, how the LNM is distributed relative to those components, and whether the observed CNM can be interpreted as material formed through dynamical compression followed by radiative cooling and thermal instability.  Resolving these issues is necessary for connecting the large scale dynamics of the infalling Draco IVC with the formation of cold atomic gas and, ultimately, molecular gas.

In this study, we investigate the H\,\textsc{i} phase structure of Draco using 21-cm data from the Dominion Radio Astrophysical Observatory H\,\textsc{i} Intermediate Galactic Latitude Survey (DHIGLS). We apply ROHSA to decompose the H\,\textsc{i} emission into spatially coherent Gaussian components, and classify the resulting components primarily according to their characteristic line widths.  We identify two kinematically distinct WNM components together with LNM and CNM components and examine their spatial and velocity distributions across the Draco region. We also compare the atomic phases with CO emission in order to investigate the relationship between the atomic and molecular components of the cloud. This paper focuses on the study of spatial and velocity structures. Detailed physical properties regarding each clump of the CNMs etc. are discussed by another paper.

The paper is organized as follows.  Section 2 describes the H\,\textsc{i} and CO data sets. Section 3 presents the ROHSA decomposition.  Section 4 describes the classification, spatial distribution, kinematics, and mass fractions of the H\,\textsc{i} phases.  Section 5 discusses the evidence for interacting WNM flows, the phase transition from the WNM through the LNM to the CNM, and a possible evolutionary scenario for cold gas and molecular cloud formation in Draco.  The conclusions are summarized in Section 6.

\section{Data}

\subsection{H\,\textsc{i}}

% Unchanged section restored verbatim from the supplied TeX.
We use HI 21-cm data from the Dominion Radio Astrophysical Observatory HI Intermediate Galactic Latitude Survey (DHIGLS; \cite{Blagrave2017}) to investigate the atomic gas structure of the Draco intermediate velocity cloud (IVC). DHIGLS combines interferometric observations from the DRAO Synthesis Telescope with single dish data from the Green Bank Telescope obtained as part of the GBT HI Intermediate Galactic Latitude Survey (GHIGLS; \cite{Martin2015}).

The angular and velocity  channel separations of the DHIGLS data are $\sim 18^{\prime\prime}$ and $\sim 0.82~{\rm km~s^{-1}}$, respectively, while the angular and velocity resolution of the dat
a are $\sim 1^{\prime}$ and $\sim 1.32~{\rm km~s^{-1}}$, respectively. Assuming a distance of 600 pc for Draco \citep{MivilleDeschenes2017}, the angular resolution corresponds to a physical scale of $\sim 0.17$ pc. The data cover a velocity range of approximately $-103 \lesssim v_{\rm LSR} \lesssim 32~{\rm km~s^{-1}}$, making them well suited for studying the spatial and kinematic structure of atomic gas associated with Draco.

\subsection{CO}

% Unchanged section restored verbatim from the supplied TeX.
\begin{figure}
 \begin{center}
 \includegraphics[width=8cm]{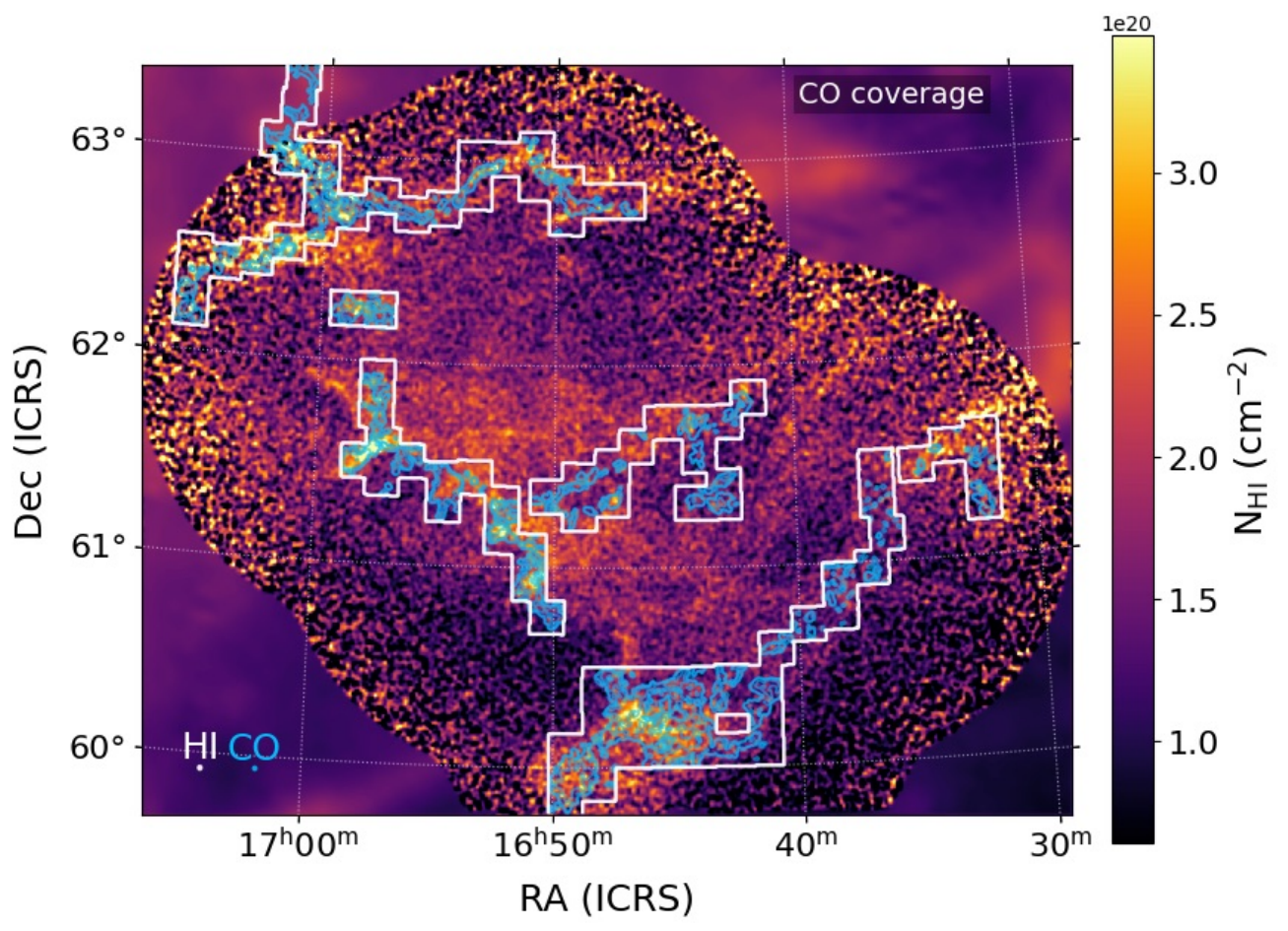}
 \end{center}
\caption{
HI column density map of the Draco cloud derived from the DHIGLS data
integrated over $-45 \lesssim v_{\rm LSR}\lesssim 5~{\rm km~s^{-1}}$.
The cyan contours show the integrated intensity of the
$^{12}$CO($J=2$--1) emission observed by \citet{BiegingKong2024},
reprojected onto the HI grid.
Contour levels correspond to the 93rd, 96th, and 99th percentiles of the
CO integrated-intensity distribution.
The white solid line indicates the original footprint of the CO
observations.\\
}
\label{fig:Draco}
\end{figure}

We use the archived deta of $^{12}$CO($J=2$--1) data of the Draco nebula obtained by \citet{BiegingKong2024}.
The observations were carried out with the Heinrich Hertz Submillimeter Telescope (HHT) of the Arizona Radio Observatory.
The data have an angular resolution of $\sim38\arcsec$, a spectral resolution of 250~kHz corresponding to a velocity resolution of $0.33~{\rm km~s^{-1}}$ at 230~GHz, and a pixel size of $10\arcsec$.
The mapped region covers all CO emission associated with structures detected by \textit{Herschel} at 250~$\mu$m with intensities exceeding $5~{\rm MJy~sr^{-1}}$, and the analyzed spectral coverage of the data cube spans $-43 \lesssim v_{\rm LSR} \lesssim -8~{\rm km~s^{-1}}$.
\citet{BiegingKong2024} also obtained spectra at a coarser resolution of 1~MHz ($\approx 1.3~{\rm km~s^{-1}}$) covering a total velocity width of $\sim 330~{\rm km~s^{-1}}$, and confirmed that no CO emission is detected at velocities outside the intermediate velocity range of the Draco nebula.
In this work we use the velocity range $-30 \lesssim v_{\rm LSR} \lesssim -12~{\rm km~s^{-1}}$, which encompasses all the CO emission from the cloud.
Figure~\ref{fig:Draco} shows the spatial distributions of the atomic
and molecular gas. The CO emission is confined to compact regions
within the H\,\textsc{i} cloud, indicating that CO bright molecular gas occupies
only a limited fraction of the overall Draco structure.

\section{Analysis: ROHSA decomposition}

To separate the H\,\textsc{i} emission into physically distinct components, we applied the ROHSA algorithm \citep{Marchal2019}.  ROHSA models each spectrum as a combination of spatially coherent Gaussian components while enforcing spatial regularization over the entire data cube.  Because the number of H\,\textsc{i} components along the line of sight toward the Draco region is not known a priori, we tested decompositions with the number of Gaussian components ranging from 2 to 10.  Based on these tests, we adopted a five components decomposition as the most appropriate representation of the observed H\,\textsc{i} spectra across the field.  The resulting Gaussian parameters were then examined in the parameter space defined by centroid velocity, line width, and peak brightness temperature.  The classification of the Gaussian components and their physical interpretation are presented in Section 4.

\begin{figure}
 \begin{center}
 \includegraphics[width=8cm]{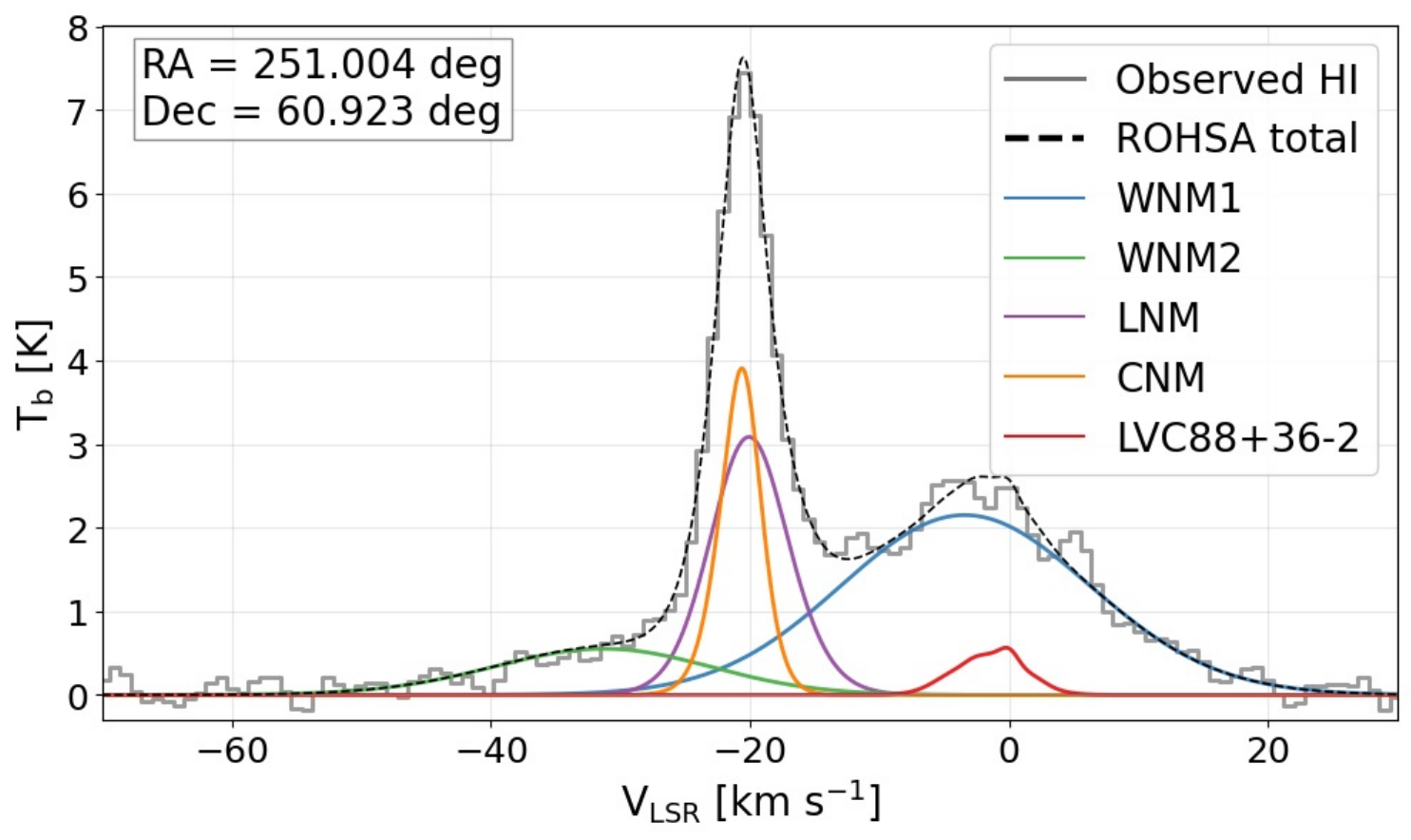}
 \end{center}
\caption{
Representative H\,\textsc{i} spectrum toward a CNM-rich position in the Draco cloud and its ROHSA Gaussian decomposition.
The histogram shows the observed H\,\textsc{i} spectrum, and the dashed curve the sum of all Gaussian components derived by ROHSA.
The individual Gaussian components, labeled in the legend, correspond to WNM1, WNM2, LNM, CNM, and the unrelated foreground cloud LVC88+36--2.
The decomposition successfully reproduces the observed line profile and illustrates the contribution of each atomic phase to the total emission.\\
}
\label{fig:spe}
\end{figure}

\begin{figure}
\begin{center}
\begin{tabular}{c}
\includegraphics[width=8cm]{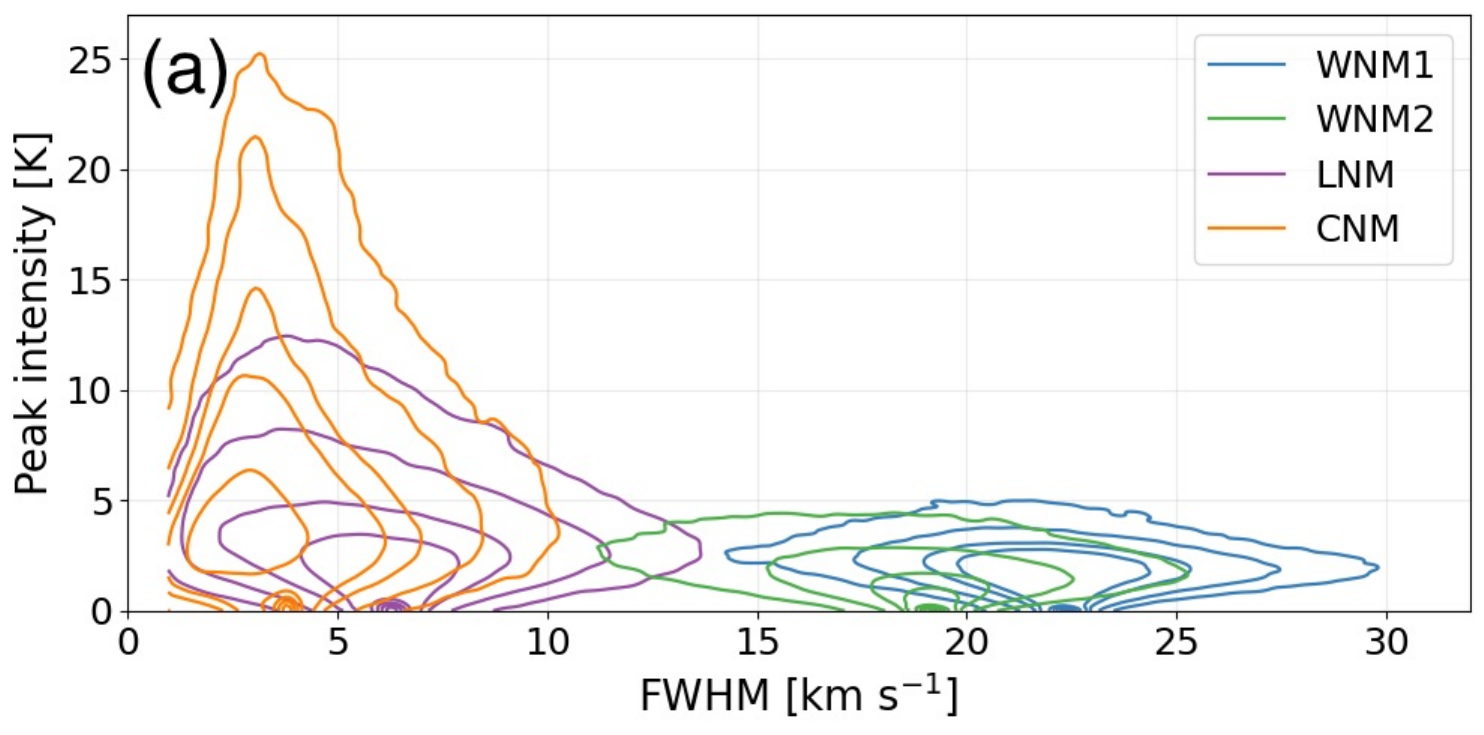}\\
\includegraphics[width=8cm]{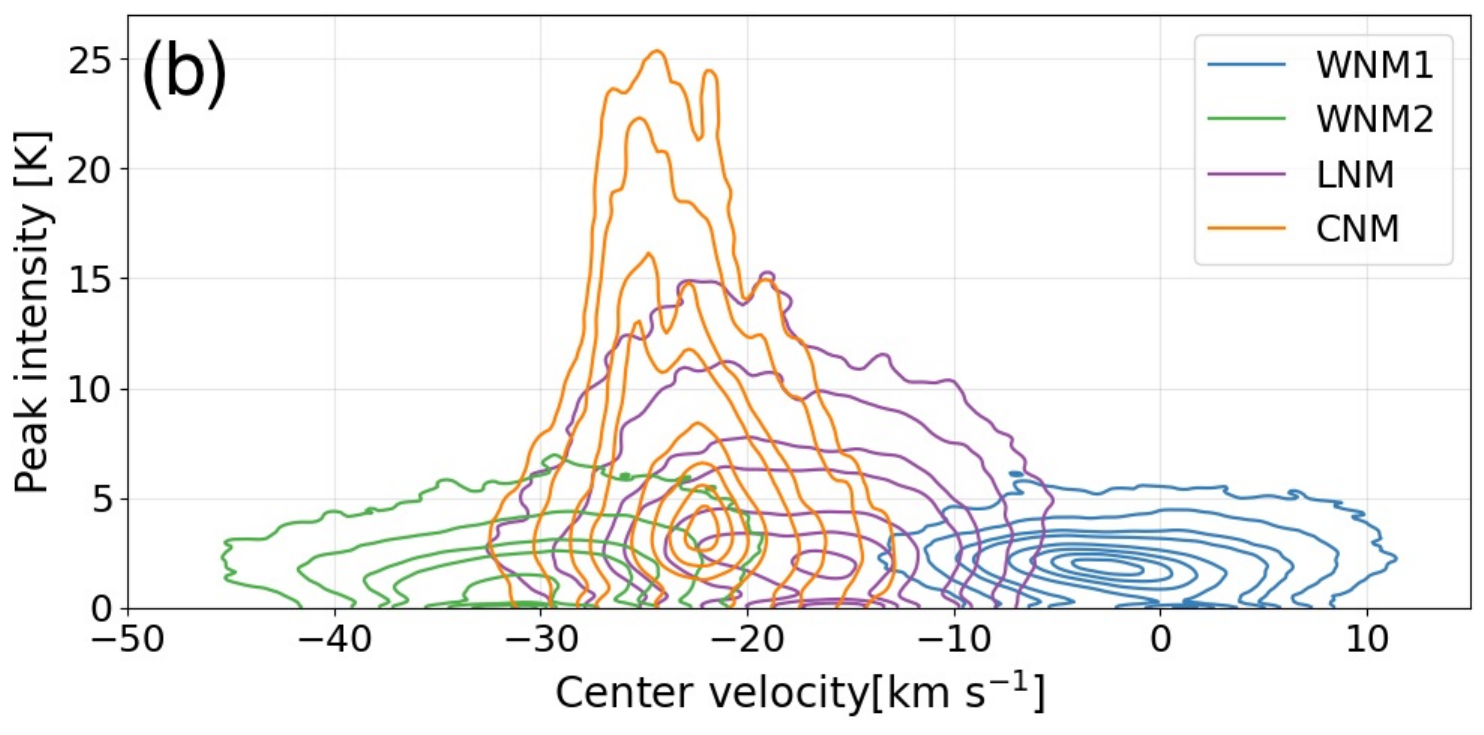}\\
\end{tabular}
\end{center}
\caption{
Distributions of the Gaussian components obtained from the ROHSA decomposition.
(a) FWHM--peak brightness temperature distribution.
(b) Centroid velocity--peak brightness temperature distribution.
The contours show the distributions of WNM1, WNM2, LNM, and CNM.
For each component, the contour levels are independently normalized to
its maximum two-dimensional KDE density and are drawn at 0.1, 1, 5, 10,
30, 50, 70, and 90\% of the maximum.\\
}
\label{fig:kde}
\end{figure}

\begin{figure*}[t]
\centering
\includegraphics[width=\textwidth]{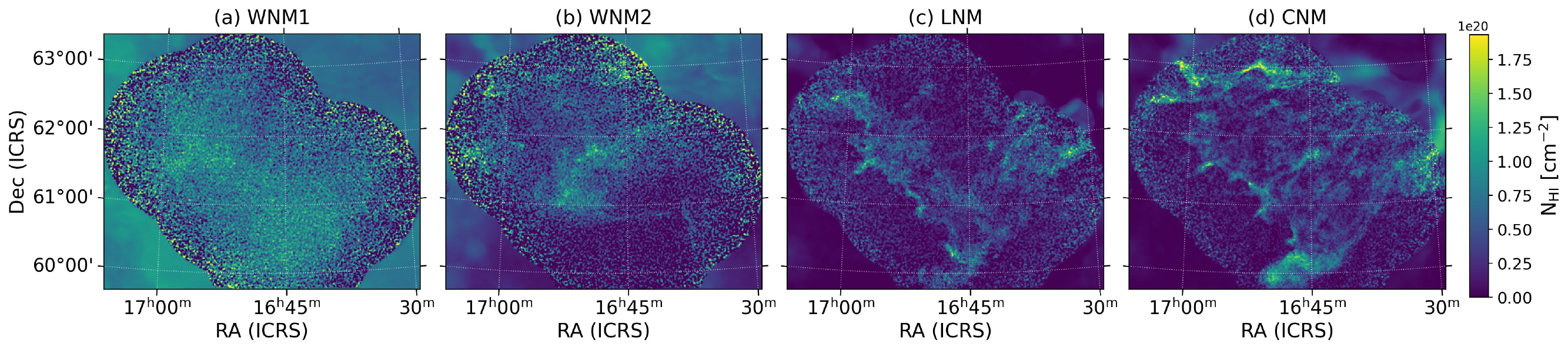}
\caption{
H\,\textsc{i} column-density maps of the four components identified from the ROHSA decomposition:
(a) WNM1 ($-24 < v_{\rm LSR}<16$ km s$^{-1}$),
(b) WNM2 ($-50 < v_{\rm LSR}<-10$ km s$^{-1}$),
(c) LNM ($-32 < v_{\rm LSR}<-10$ km s$^{-1}$), and
(d) CNM ($-32 < v_{\rm LSR}<-12$ km s$^{-1}$).
A common color scale is adopted for all panels to facilitate direct comparison of their spatial distributions.\\
}
\label{fig:ALL}
\end{figure*}

\section{Results}

Figure~\ref{fig:spe} shows a representative H\,\textsc{i} spectrum toward a CNM rich position together with its Gaussian decomposition.  The observed spectrum is well reproduced by the sum of the five Gaussian components, illustrating the contribution of each component to the total H\,\textsc{i} emission.  Based on its morphology and velocity, one of the five components is identified as the nearby low-velocity cloud LVC88+36$-$2 \citep{Wennmacher1992}.  Previous studies have shown that LVC88+36$-$2 is an unrelated foreground cloud that overlaps the Draco nebula only in projection on the plane of the sky.  In the following analysis, this foreground component is treated separately, and we focus on the remaining four Gaussian components associated with the Draco nebula.

\subsection{Gaussian components and phase separation}

We first examined the distributions of the Gaussian components obtained from the ROHSA decomposition.  As shown in Figure~\ref{fig:kde}(a), the components are broadly separated into two groups in the FWHM -- peak temperature plane.  The broad line group has a characteristic FWHM of $\sim$20 km s$^{-1}$, whereas the narrow line group shows two characteristic peaks at $\sim$3 and $\sim$6 km s$^{-1}$.  These groups are categorized by different H\,\textsc{i} phases primarily on the basis of their characteristic line widths, following the linewidth regimes from a Gaussian decomposition of the all-sky HI4PI data \citep{Kalberla2018}, and that from GALFA-HI data \citep{Matsuzuki2026}.  The broad components have relatively low peak brightness temperatures, and are identified with the warm neutral medium (WNM).  The narrower components have higher peak brightness temperatures, and are interpreted as the cold and thermally unstable neutral phases.  Within this group, the components with characteristic FWHM values of $\sim$6 km s$^{-1}$ generally have lower peak brightness temperatures than those with FWHM values of $\sim$3 km s$^{-1}$.  We therefore associate the $\sim$3 km s$^{-1}$ components with the Cold Neutral Medium (CNM), and the $\sim$6 km s$^{-1}$ components with the Lukewarm Neutral Medium (LNM), which represents the thermally unstable phase between the WNM and CNM.  However, the CNM and LNM distributions are not clearly separated in the FWHM--peak temperature plane.  This may partly reflect the relatively coarse velocity resolution of the DHIGLS data, 1.32 km s$^{-1}$, which can limit the ability to distinguish narrow CNM components from somewhat broader LNM components.  To examine the effect of velocity resolution on the phase separation, we performed an additional test using GALFA-HI data with an original velocity resolution of 0.184 km s$^{-1}$.  The data were progressively smoothed along the velocity axis to coarser resolutions, and the ROHSA decomposition was repeated for each smoothed data set.  We found that the CNM and LNM distributions in the FWHM--peak temperature plane increasingly overlap once the velocity resolution becomes poorer than $\sim$0.9 km s$^{-1}$ (See Appendix A).  This test suggests that the 1.32 km s$^{-1}$ velocity resolution of the DHIGLS data may not be sufficient to fully separate the CNM and LNM populations.  Despite this limitation, the narrow line components still exhibit two characteristic linewidth peaks near $\sim$3 and $\sim$6 km s$^{-1}$.  In the following analysis, we identify the components near 3 km s$^{-1}$ as the CNM and those near 6 km s$^{-1}$ as the LNM, while keeping in mind that some overlap or misclassification between these two phases may remain.  The velocity -- peak temperature distribution shown in Figure~\ref{fig:kde}(b) further reveals that the two WNM components are separated in velocity space.  The component centered near a local standard of rest velocity of -3 km s$^{-1}$, hereafter WNM1, is clearly separated from the component concentrated around approximately $-$30 km s$^{-1}$, hereafter WNM2. Both the LNM and CNM are distributed at intermediate velocities between the two WNM components.

\begin{figure*}[t]
\centering
\includegraphics[width=0.75\textwidth]{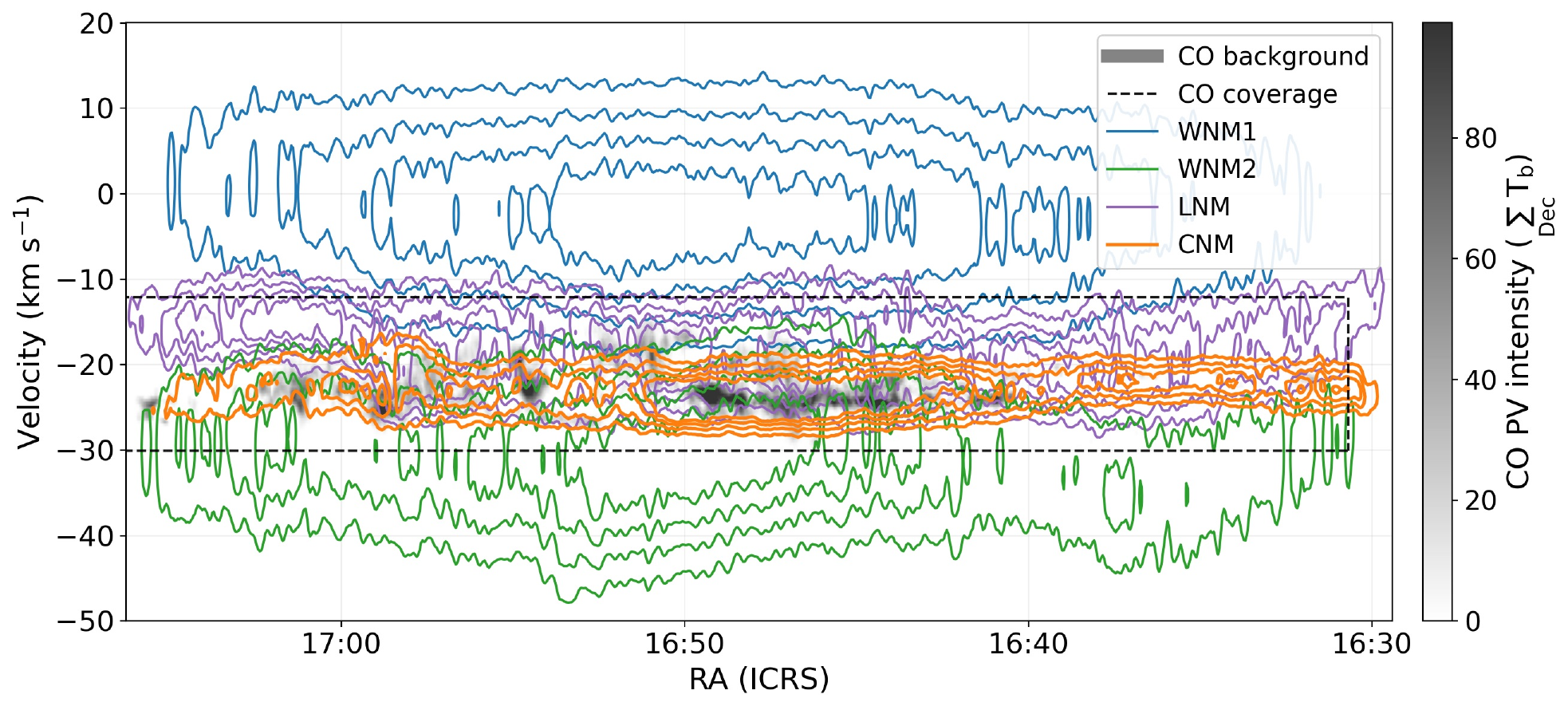}\par
\medskip
\includegraphics[width=0.75\textwidth]{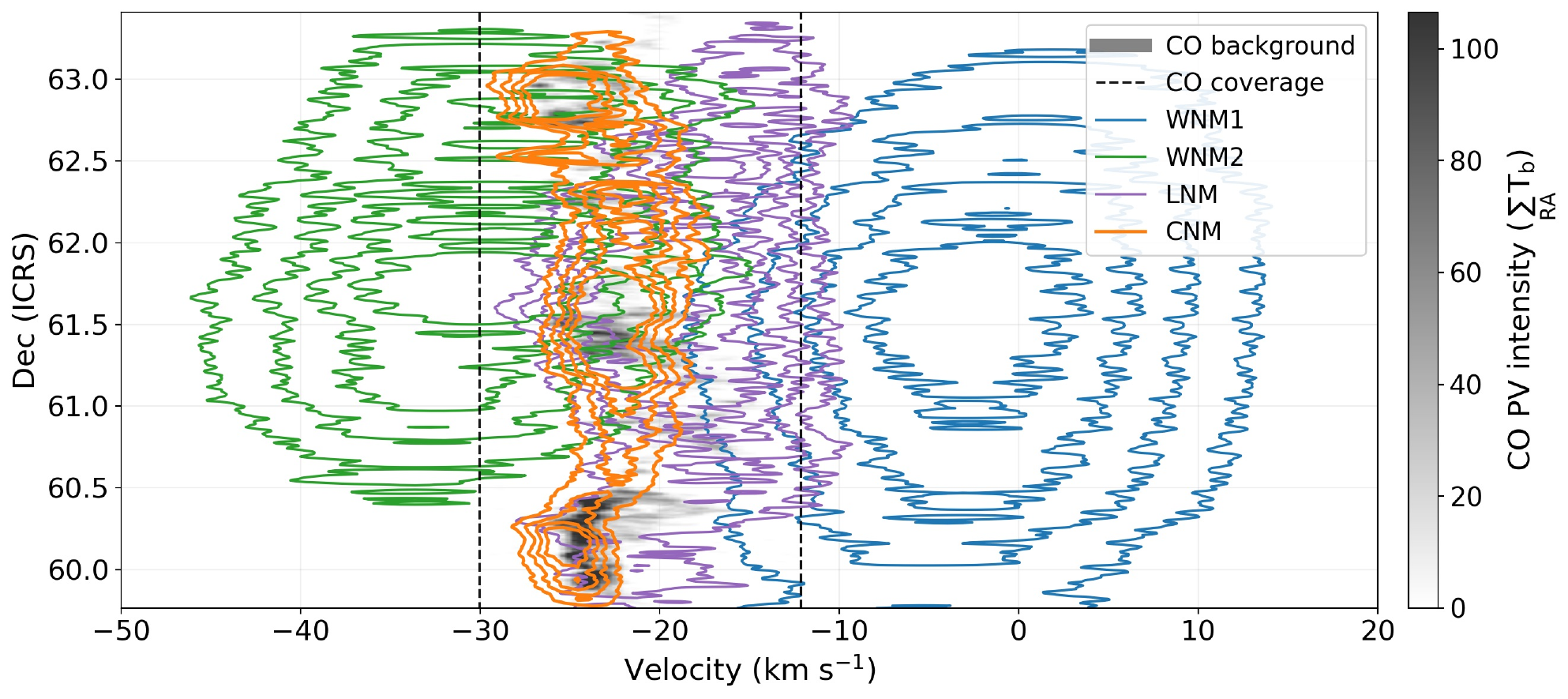}
\caption{
Phase-separated position--velocity diagrams of the H\,\textsc{i} gas in Draco.
(a) Velocity--RA diagram integrated along the Declination axis.
(b) Declination--velocity diagram integrated along the Right Ascension axis.
Colored contours indicate the distributions of WNM1 (blue), WNM2 (green),
LNM (purple), and CNM (orange).
Contour levels are set at 30\%, 50\%, 70\%, and 90\%
of the 99th-percentile intensity of each component in each panel.
The grayscale image represents the CO position--velocity intensity,
and the dashed outline indicates the spatial and velocity extent of the
CO observational coverage.\\
}
\label{fig:pv}
\end{figure*}

\subsection{Spatial distributions of the H\,\textsc{i} phases}

% Unchanged section restored verbatim from the supplied TeX.
Figure~\ref{fig:ALL} presents the column-density distributions of the four H\,\textsc{i} phases.
The column density of each phase was derived from
\begin{equation}
N_{\rm HI}
=
1.822\times10^{18}
\int T_{\rm B}\,dv
\quad
{\rm cm^{-2}},
\end{equation}
assuming optically thin H\,\textsc{i} emission.

Because no opacity correction is applied, the derived CNM column
density, and mass may be underestimated if the cold H\,\textsc{i} emission is
optically thick.
This uncertainty is up to $\sim$30 \% at high Galactic latitude \citep{Matsuzuki2026}.
Both WNM1 and WNM2 exhibit diffuse, and extended distributions across the Draco region. WNM1 shows a relatively smooth morphology, whereas WNM2 displays more pronounced spatial substructures, particularly around the central part of the cloud.

The LNM forms an interconnected network of filamentary structures extending over much of the cloud. The most prominent feature is a gently curved filamentary complex spanning approximately
$\mathrm{RA}\simeq16^{\rm h}45^{\rm m}$--$17^{\rm h}00^{\rm m}$ and
$\mathrm{Dec}\simeq60^\circ00^\prime$--$63^\circ00^\prime$.

In contrast, the CNM is distributed in localized bright structures. Two prominent filamentary complexes are identified. The northern complex is characterized by several elongated filaments extending approximately east--west around
$\mathrm{Dec}\simeq62^\circ40^\prime$--$63^\circ00^\prime$.
The southern complex is located around
$\mathrm{RA}\simeq16^{\rm h}45^{\rm m}$ and
$\mathrm{Dec}\simeq59^\circ50^\prime$--$60^\circ20^\prime$,
where a bright condensation is accompanied by several short filamentary structures. Additional compact condensations are present near the western edge of the cloud and in the central region. Compared with the LNM, the CNM exhibits a more fragmented morphology. Many CNM structures are spatially associated with the LNM network, although strong CNM emission is also detected in regions where the LNM contribution is relatively weak.

\subsection{Mass fractions of the H\,\textsc{i} phases}

% Interaction regionだけの比を入れたい。計算して。

The H\,\textsc{i} mass of each component was calculated as

\begin{equation}
M_{\rm HI}
=
m_{\rm H}D^2
\sum_i N_{{\rm HI},i}\Omega_{\rm pix},
\end{equation}

where $D=600$~pc, $\Omega_{\rm pix}$ is the solid angle of one pixel,
and the sum was evaluated over the analyzed region. The quoted masses
represent atomic hydrogen only and do not include a correction for
helium.

\begin{equation}
M_{\rm HI,tot}
\simeq
1687\,M_\odot .
\end{equation}

The individual phase masses are
$M_{\rm WNM1}\simeq693\,M_\odot$,
$M_{\rm WNM2}\simeq395\,M_\odot$,
$M_{\rm LNM}\simeq282\,M_\odot$,
and
$M_{\rm CNM}\simeq317\,M_\odot$.

The resulting mass fractions are

\begin{equation}
{\rm WNM1:WNM2:LNM:CNM}
=
41:23:17:19.
\end{equation}
%？？？？
Gaussian decompositions of H\,\textsc{i} emission-line data reveal large
regional variations. \citet{Marchal2021} obtained
CNM:LNM:WNM $=8:28:64$ in the NEP region,
\citet{Taank2022} found
CNM:LNM:WNM $=50:21:27$ for the LVC on the NCPL,
and \citet{Matsuzuki2026} reported
CNM:LNM:WNM $=43:47:10$ for the molecular-cloud-forming atomic cloud
HLCG~92--35.

The Draco cloud presents a contrasting case, with
CNM:LNM:WNM $=19:17:64$.
The WNM still accounts for about two thirds of the total H\,\textsc{i} mass,
indicating that a substantial reservoir of warm atomic gas remains.
Even in taking into account the optically thick case of the CNM, CNM:LNM:WNM mass ratio = 23:16:61, which is still WNM is substantial reservoir.
This is in marked contrast to HLCG~92--35, where the WNM constitutes
only $\sim 10\%$ of the total H\,\textsc{i} mass.
At the same time, the LNM fraction in Draco is lower than the CNM
fraction, whereas the studies cited above generally find
LNM $\gtrsim$ CNM.

\section{Discussion}

\subsection{Evidence for colliding WNM flows}

The presence of two WNMs separated by velocity components indicates that distinct warm atomic gas components coexist along the line of sight toward the Draco nebula. This separation is clearly confirmed in the position--velocity diagrams shown in Figure~\ref{fig:pv}. WNM1 is distributed predominantly around a local velocity of $\sim -3$ km s$^{-1}$, whereas WNM2 occupies substantially more negative velocities at $\sim -30$ km s$^{-1}$. Between these two components, the LNM extends continuously over a broad spatial range in both the velocity--RA and Dec--velocity diagrams, forming a bridge in velocity space. Such an intermediate-velocity component is qualitatively consistent with the kinematic structure expected from interacting atomic flows. Numerical simulations of converging WNM flow show that gas entering an interaction region is compressed and decelerated, producing a shock-compressed layer whose bulk velocity lies between those of the two inflows \citep{Kobayashi2020}. For flows with comparable momentum fluxes, the velocity of the interaction layer approaches the mean velocity of the inflowing components, although the exact value depends on their densities, velocities, and interaction geometry. The occurrence of the LNM at velocities intermediate between WNM1 and WNM2 is therefore consistent with dynamically processed gas produced through an interaction between the two warm components \citep{Audit2005,Micic2013,Kobayashi2020}. The CNM is concentrated within a narrower part of the intermediate-velocity range, particularly around $-20$ to $-27$ km s$^{-1}$. The CO emission is also concentrated within approximately the same velocity interval, and is closely associated with CNM-rich regions in the position--velocity diagrams. This nested kinematic structure is consistent with a dynamically compressed multiphase interaction layer, with the LNM bridging the two WNM components, and the CNM and CO confined to a narrower range of intermediate velocities. However, we emphasize that this ordering in velocity space alone does not establish an evolutionary sequence among the thermal phases.

An additional feature of the intermediate-velocity gas is the absence of a clearly separable WNM component in the present ROHSA decomposition. Although the LNM and CNM are prominent between WNM1 and WNM2, no comparably strong broad WNM component is independently recovered at these velocities. Because WNM emission is intrinsically broad, a weak intermediate-velocity WNM component could have a low peak brightness temperature even if its integrated intensity were non-negligible. Such broad, low-amplitude emission may be difficult to isolate as an independent Gaussian component and may instead be partly incorporated into the fitted wings of WNM1 or WNM2. We therefore cannot exclude weak warm gas at intermediate velocities. The present decomposition shows only that the detectable velocity bridge is dominated by LNM and CNM emission.

We interpret WNM1 as predominantly ambient atomic gas associated with the local Galactic environment, whereas WNM2 represents the higher-velocity component associated with the infalling Draco IVC. This interpretation is consistent with previous studies in which Draco has been described as an infalling cloud interacting with diffuse Galactic gas \citep{MivilleDeschenes2017,Schneider2024}. Independent support for this interaction is provided by the detection of the 158 $\mu$m [C\,\textsc{ii}] line toward Draco by \citet{Schneider2024}. They detected [C\,\textsc{ii}] emission at intermediate velocities in four of five observed positions, whereas no corresponding emission was detected at local or higher velocities. The [C\,\textsc{ii}] emission overlaps the velocity interval in which the LNM, CNM, and CO are prominent. \citet{Schneider2024} found that the observed [C\,\textsc{ii}] intensities are difficult to reproduce under the weak far-ultraviolet radiation field of Draco using conventional PDR heating alone, and proposed an additional contribution from mechanical heating associated with shocks and H\,\textsc{i} cloud interactions. Therefore, the [C\,\textsc{ii}] observations provide independent evidence that mechanical energy is being dissipated in the intermediate-velocity gas. Taken together, the two velocity-separated WNM components, the LNM velocity bridge, the concentration of CNM and CO within the intermediate-velocity interval, and the [C\,\textsc{ii}] evidence for mechanical heating support a picture in which the infalling Draco IVC interacts with more local Galactic WNM and produces a dynamically compressed multiphase layer.

\subsection{Phase transition from the WNM to the CNM}

\subsubsection{Observational signatures of the WNM-to-CNM transition}

We next examine the thermal evolution of the atomic gas within this interaction layer.
The velocity distributions of the LNM and CNM provide an important constraint. Although both phases lie predominantly at velocities between WNM1 and WNM2, their centroid velocities are not identical. As shown in Figure~\ref{fig:pv}, the centroid velocity of the CNM is systematically shifted by a few km s$^{-1}$ toward WNM2 relative to that of the LNM. Thus, the CNM lies within the overall velocity range occupied by the LNM but is displaced toward the velocity of the infalling component. This systematic offset suggests that the LNM and CNM do not simply represent different thermal states of gas sharing an identical bulk velocity. Rather, the transition toward the cold phase appears to occur while the gas is also evolving dynamically. In a purely local thermal transition occurring at fixed bulk velocity, such a systematic centroid displacement would not necessarily be expected. The observed offset therefore favors a picture in which cooling and condensation occur within a dynamically evolving compressed layer in which momentum exchange accompanies thermal phase conversion. The CNM displacement toward WNM2 may indicate that the cold gas retains a stronger kinematic imprint of the infalling component than does the LNM. This interpretation is not unique, because projection effects, turbulent motions, and uncertainties associated with the decomposition of overlapping spectral components may also contribute to the observed velocity differences. Nevertheless, when combined with the independent evidence for interaction between WNM1 and WNM2, the CNM--LNM velocity offset provides an additional observational constraint on the phase-transition process.

The mass fractions of the phases also indicate substantial thermal processing. The LNM and CNM together account for $\sim36\%$ of the total H\,\textsc{i} mass in the analyzed region. This value should not be interpreted directly as the conversion efficiency from WNM to CNM because the initial phase composition, evolutionary stage, and possible line-of-sight contamination are unknown. Nevertheless, when combined with the independent evidence for the interaction between WNM1 and WNM2, the CNM--LNM velocity offset provides an additional observational constraint on the phase-transition process, and strengthens the interpretation that the CNM is formed through dynamically driven cooling and condensation of compressed atomic gas.

\subsubsection{Cooling, dynamics, and thermal conduction}

The observed phase structure is qualitatively consistent with theoretical models of converging atomic flows, in which compression of initially warm gas increases its density, and shortens its radiative cooling time, thereby allowing it to enter the thermally unstable regime and, under favorable conditions, condense toward the CNM \citep{Hennebelle1999,Audit2005,Saury2014,Kobayashi2020}. Sustained quasi-isothermal compression that increases the mean WNM density by a factor of $\sim2$--4 can result in substantial CNM formation, with representative models producing cold-gas fractions of $\sim40\%$ within $\sim1$ Myr \citep{Saury2014}. The efficiency of this transition depends on the competition between radiative cooling and dynamical evolution. Condensation can proceed efficiently when the cooling time is shorter than the relevant dynamical time \citep{Hennebelle1999,Saury2014}. Conversely, rapid compression and decompression can inhibit condensation when dynamical motions act faster than radiative cooling \citep{Audit2005,Saury2014}.

Thermal conduction introduces an additional scale dependence. Perturbations smaller than the Field length are smoothed by thermal conduction, and are stabilized against thermal condensation, whereas larger-scale perturbations can grow through radiative cooling \citep{Field1965,Koyama2004}. In strongly dynamical condensation fronts, the relative importance of conduction decreases as the mass flux increases, and the thickness of the transition layer can instead be controlled by the cooling length \citep{Iwasaki2012}. The spatial resolution of the present H\,\textsc{i} observations is $\sim0.4$ pc, substantially larger than the characteristic Field length expected for thermally bistable atomic gas. The conductive interfaces and the smallest scales on which thermal conduction regulates phase boundaries are therefore unresolved. Our analysis consequently probes the phase structure on sub-parsec to parsec scales rather than the internal structure of individual conductive transition fronts.

\subsubsection{Comparison with colliding-flow simulations}

Numerical simulations of colliding atomic flows provide further insight into the thermal evolution of such compressed gas. \citet{Micic2013} showed that warm atomic gas entering a converging-flow interaction region is strongly modified by compression and radiative cooling, and that the resulting velocity distribution is closely coupled to the thermal evolution of the shocked material. Their simulations demonstrate that cooling physics has a strong influence on the large-scale velocity structure of the gas formed in the interaction region, illustrating the close connection between kinematics and thermal phase evolution.

\citet{Kobayashi2020} similarly found that shock-compressed WNM rapidly develops into a multiphase layer containing stable WNM, thermally unstable gas, and CNM, with the stable WNM accounting for only a relatively small fraction of the mass in the compressed layer. The partition between thermally unstable gas and CNM depends strongly on the structure of the upstream WNM. When the inflowing gas contains large density fluctuations, the shock fronts become highly distorted and generate stronger post-shock turbulence. This turbulence reduces the efficiency of condensation and leaves a comparatively large fraction of the gas in the thermally unstable phase, with the CNM mass fraction remaining around 45\%. For smoother inflows, the shocks remain more nearly planar, kinetic energy is dissipated more efficiently, and the CNM fraction can increase to $\sim70\%$. These calculations therefore demonstrate that a substantial thermally unstable component can persist in a dynamically compressed layer even while CNM formation proceeds efficiently. This behavior provides a natural framework for interpreting the phase balance observed in Draco. The observations therefore do not require the interaction layer to evolve uniformly toward the CNM; rather, different parts of the layer may be at different thermal stages depending on local compression, turbulence, and shock geometry.

The comparison by \citet{Schneider2025} with numerical cloud-formation models provides additional support for this interpretation. They found that the observed column-density distribution in Draco is better reproduced when both cold and thermally unstable atomic gas are included than when the CNM alone is considered. This result is qualitatively consistent with the substantial LNM component identified in our decomposition, and supports a picture in which thermally unstable gas remains an important constituent of the cloud during its evolution toward colder and molecular phases.

\citet{Kobayashi2022} further showed that the turbulence generated in multiphase converging-flow layers becomes predominantly solenoidal despite the compressive nature of the large-scale driving. They also found that individual CNM structures maintain relatively small internal velocity dispersions, while their bulk motions are dynamically coupled to turbulence in the surrounding WNM and thermally unstable gas. This provides a plausible framework for interpreting the kinematic relationship between the LNM and CNM in Draco described in Section~5.2.1. The cold structures can remain dynamically associated with the surrounding unstable gas without requiring identical centroid velocities or comparable internal line widths. The narrower characteristic line width of the CNM relative to the LNM is qualitatively consistent with such a multiphase turbulent structure. We caution, however, that the thermally unstable gas defined in the simulations is not necessarily identical to the LNM identified here from the H\,\textsc{i} line-width decomposition.

The relative amounts of LNM and CNM provide an additional constraint on the thermal evolution of the interaction layer. In the present decomposition, the LNM and CNM account for $\sim17\%$ and 19\% of the total H\,\textsc{i} mass, respectively, corresponding to an LNM/CNM mass ratio of about 0.9. Thus, the thermally unstable gas does not dominate over the cold phase in the analyzed region. As discussed above, converging-flow calculations show that sustained compression can drive a substantial fraction of thermally unstable gas into the CNM, whereas stronger post-shock turbulence can maintain a larger unstable fraction \citep{Saury2014,Kobayashi2020}. The near-unity LNM/CNM ratio in Draco therefore suggests that condensation into the CNM has already proceeded substantially, while the remaining LNM indicates that the conversion is still incomplete. This phase balance is consistent with a dynamically evolving interaction layer in which radiative cooling and turbulent disruption act simultaneously.

Overall, the observations of Draco are consistent with a dynamically driven WNM--LNM--CNM transition within the interaction layer. The LNM is interpreted predominantly as compressed and cooling atomic gas, whereas the CNM traces regions in which condensation has progressed further. This interpretation is statistical rather than a requirement that every gas parcel follow a monotonic thermal or velocity sequence.

\begin{figure}
 \begin{center}
 \includegraphics[width=8cm]{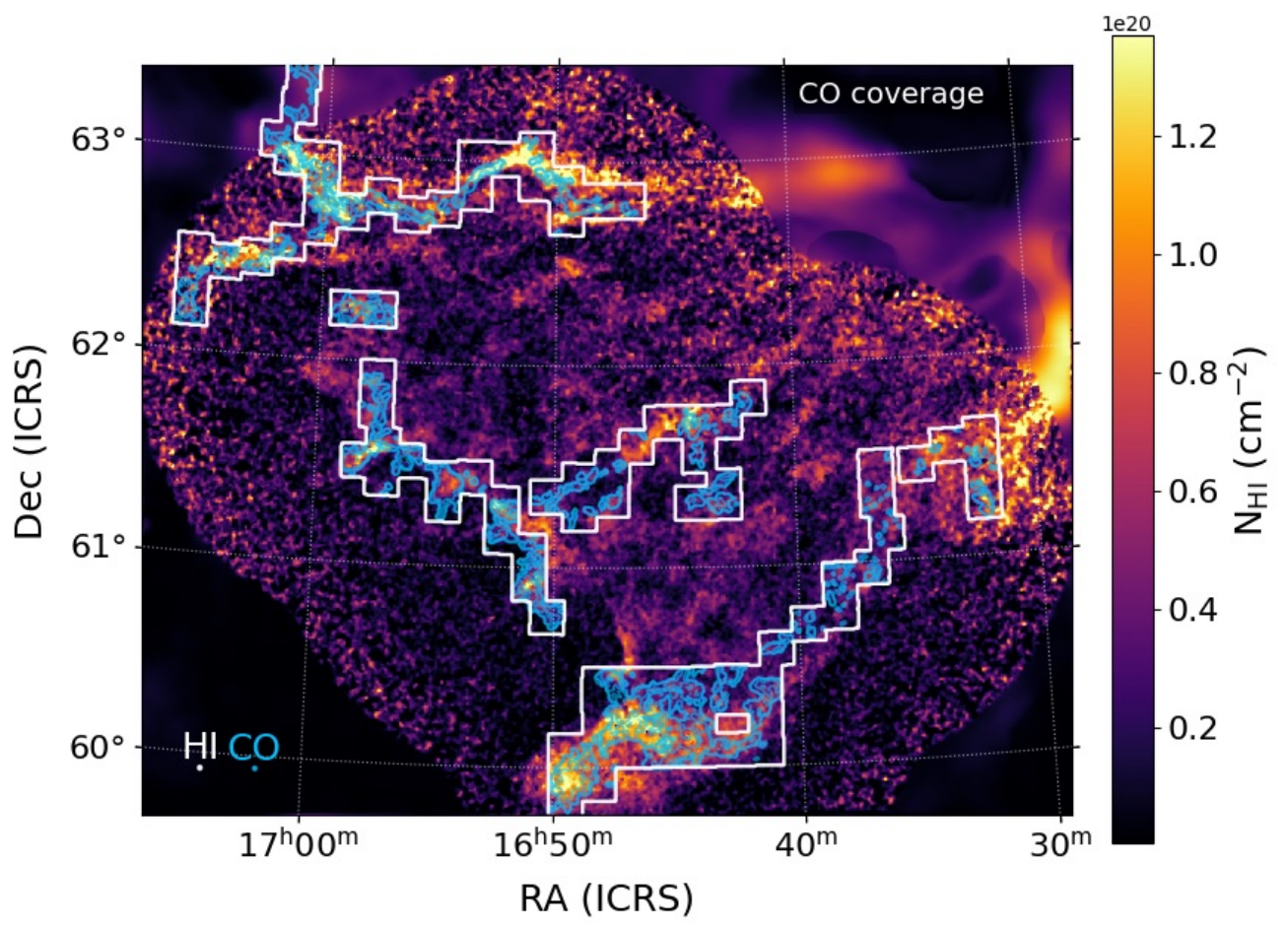}
 \end{center}
\caption{
Comparison of the spatial distributions of the CNM identified by the
ROHSA decomposition and the $^{12}$CO($J=2$--1) emission.
The background image shows the CNM H\,\textsc{i} column density, while the cyan
contours indicate the integrated CO intensity reprojected onto the H\,\textsc{i}
grid. Contour levels correspond to the 93rd, 96th, and 99th
percentiles of the CO integrated-intensity distribution.
The white solid line marks the original footprint of the CO
observations. The H\,\textsc{i} and CO beam sizes are shown in the lower-left
corner.\\
}
\label{fig:HI_CO}
\end{figure}

\begin{figure*}[t]
\centering
\includegraphics[width=\textwidth]{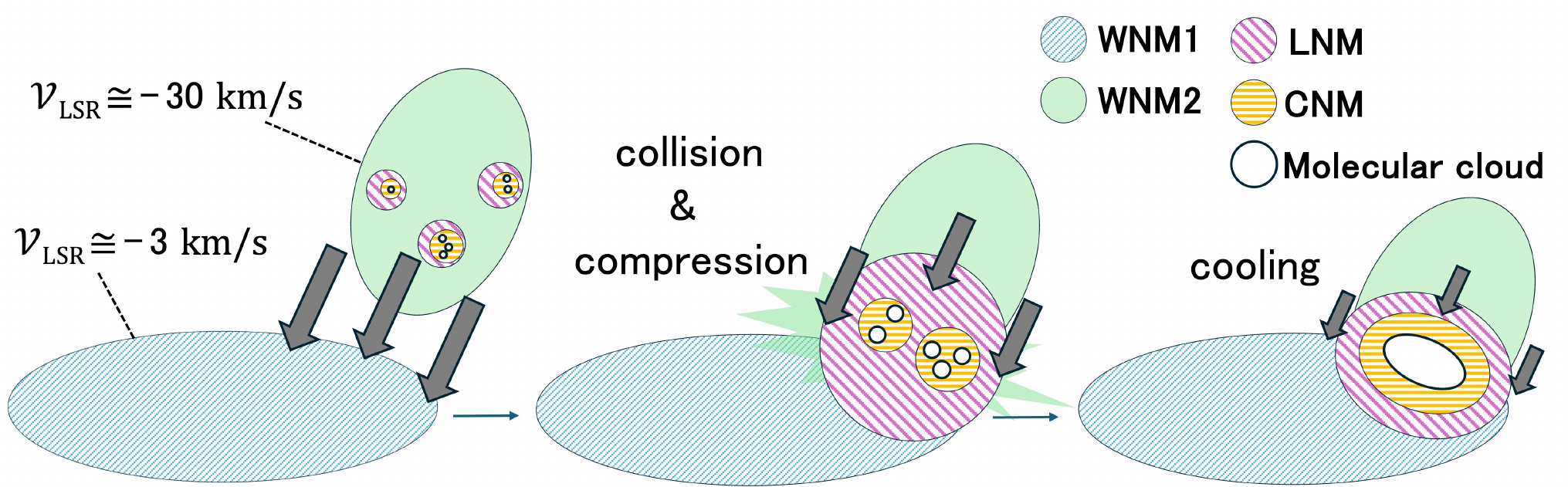}
\caption{
Schematic illustration of the proposed molecular-cloud formation scenario in Draco.
An infalling warm neutral medium component (WNM2) collides with ambient warm atomic gas (WNM1) near the Galactic disk--halo interface, producing a compressed layer of lukewarm neutral medium (LNM).
Subsequent radiative cooling leads to the formation of cold neutral medium (CNM) structures, within which molecular gas is eventually formed.
The characteristic velocity difference between the two WNM components is approximately $30~{\rm km~s^{-1}}$, while the compressed intermediate velocity gas is observed near $v_{\rm LSR}\simeq-20~{\rm km~s^{-1}}$.\\
}
\label{fig:scenario}
\end{figure*}

\subsection{Implications and proposed scenario for CNM and molecular-cloud formation in Draco}

Figure~\ref{fig:scenario} summarizes the evolutionary scenario implied by the results discussed above.

\subsubsection{Initial conditions and pre-existing dense structures}

In the first stage, illustrated in the left panel of Figure~\ref{fig:scenario}, WNM1 represents the local ambient Galactic H\,\textsc{i}, whereas WNM2 is associated with the infalling Draco IVC. Their line-of-sight velocity separation is $\sim30$ km s$^{-1}$. The velocity component normal to the actual interaction surface is unknown because the three-dimensional geometry is unconstrained, and the observed velocity difference should therefore not be interpreted directly as the collision velocity. We do not assume that the infalling component was initially spatially uniform or composed purely of WNM. Before the present interaction, WNM2 may already have contained density inhomogeneities, CNM condensations, and small amounts of molecular or CO-dark H$_{2}$ gas. Ultraviolet absorption-line observations toward background sources have detected H$_{2}$ in a number of Galactic intermediate-velocity clouds, demonstrating that molecular material can already be present within IVCs before their interaction with the Galactic disk \citep{Richter2003}. Such a multiphase initial condition is therefore physically plausible, and avoids requiring all of the present molecular material in Draco to have formed from diffuse WNM during a single interaction episode. This possibility is particularly relevant to the more massive dense structures. \citet{MivilleDeschenes2017} identified condensations with masses extending to approximately 20 $M_{\odot}$, together with a large population of subsolar-mass structures. The characteristic H\,\textsc{i} column densities measured in this study are $\sim1.5 \times 10^{20}$ cm$^{-2}$ for WNM1 and 0.5--1.2 $\times 10^{20}$ cm$^{-2}$ for WNM2. If both components contribute material to the interaction, their combined surface density is only a few $M_{\odot}$ pc$^{-2}$. Even at complete conversion efficiency, assembling a 20 $M_{\odot}$ condensation would require collecting material over several square parsecs, corresponding to a characteristic radius of order 1--2 pc. The present column-density measurements therefore do not demonstrate that the most massive condensations were assembled entirely from diffuse WNM during the current interaction. Moreover, because neither the volume density of the inflowing WNM nor the velocity component normal to the interaction surface is known, the corresponding mass-assembly timescale cannot be determined directly. However, for subsolar-mass structures, the required collection scales are only a few tenths of a parsec, comparable to the sizes of cold structures observed in Draco. Some low-mass CNM or molecular condensations could therefore plausibly have formed during the present interaction, whereas more massive structures may contain pre-existing dense material that has subsequently been compressed and grown. The observations thus allow a mass-dependent formation history.

\subsubsection{Growth of cold structures in the interaction layer}

In the second stage, illustrated in the middle panel of Figure~\ref{fig:scenario}, the CNM formed through the phase conversion discussed in Section~5.2 becomes an important reservoir for subsequent cloud growth. At the same time, any dense structures already present in the infalling material are further processed by the interaction. Pre-existing cold condensations entering the interaction region may be compressed further, acquire additional shielding, and accrete newly cooled atomic gas from their surroundings. The interaction can therefore contribute to cloud assembly in two complementary ways. It can produce new CNM from initially warmer gas, while also increasing the mass of dense structures that existed before the present encounter. This distinction is particularly useful in light of the mass budget argument presented in Section~5.3.1. The numerous low-mass condensations in Draco can plausibly be formed, at least in part, from material processed within the present interaction layer, whereas the most massive structures are more naturally explained if some pre-existing dense material was already present, and subsequently grew through compression and accretion. The present interaction therefore need not create all cold structures from initially diffuse WNM in a single event.

Such a mixed formation history is also consistent with the highly inhomogeneous nature of the interaction layer expected from converging-flow models. Different regions can experience different degrees of compression, cooling, and turbulent disruption, so newly formed CNM, growing pre-existing condensations, and more weakly processed unstable gas can coexist within the same cloud. The current interaction thus acts not only as a mechanism for generating cold atomic gas, but also as a driver of the hierarchical assembly of dense structures in Draco.

\subsubsection{From CNM to molecular gas}

In the third stage, illustrated in the right panel of Figure~\ref{fig:scenario}, the densest CNM structures evolve toward molecular conditions. Once sufficient column density and shielding are achieved, H$_{2}$ formation becomes efficient, and CO becomes abundant in the densest regions. Previous studies have examined this later stage in detail. \citet{MivilleDeschenes2017} identified numerous dense dust structures at the compressed Draco interface and interpreted their morphology in terms of dynamical interaction and Rayleigh--Taylor instability. They also inferred a characteristic turbulent dissipation scale of $\sim0.1$ pc, suggesting that kinetic energy injected on larger scales is transferred toward smaller structures within the cloud. \citet{BiegingKong2024} showed that the CO emission is highly clumpy, and that large-scale relative motions among molecular structures coexist with much smaller internal motions. They further inferred evolutionary timescales of a few $10^{5}$ yr for individual CO structures. These relatively short timescales characterize the local evolution of already dense molecular material, and should not be interpreted as the time required to assemble the entire molecular mass of Draco from diffuse WNM. \citet{Schneider2025} further argued that Draco is undergoing a non-equilibrium H\,\textsc{i}-to-H$_{2}$ transition at an early stage of molecular-cloud evolution. For representative conditions, they estimated a turbulent mixing timescale of $\sim1.87$ Myr and an H$_{2}$ chemical formation timescale of $\sim3.3$ Myr. The fact that the dynamical timescale is shorter than the chemical timescale implies that the gas can be dynamically rearranged before reaching chemical equilibrium, emphasizing the time-dependent nature of the atomic-to-molecular transition.

As discussed in Section~5.3.1, the molecular component may contain structures with different evolutionary histories. Some may have formed recently from CNM produced in the interaction layer, whereas others may reflect the subsequent chemical evolution and growth of material that was already dense before the present interaction. The short chemical or evolutionary timescales inferred for individual molecular structures should therefore not be interpreted as the time required to assemble the entire molecular mass of Draco from diffuse WNM. Molecular chemistry can proceed rapidly once sufficiently dense and shielded gas exists, whereas the assembly of the required cold atomic material may occur on a different timescale.

The proposed scenario can therefore be summarized schematically as

WNM $\rightarrow$ compressed atomic gas $\rightarrow$ LNM $\rightarrow$ CNM $\rightarrow$ molecular gas,

while allowing pre-existing cold and molecular structures to participate in the present interaction. The sequence describes the statistical evolution of the gas rather than a unique trajectory followed by every parcel. We therefore propose that CNM formation within the compressed interaction layer is the key process linking the dynamical interaction in Draco to its subsequent molecular-cloud evolution.

\section{Summary}
\label{sec:summary}

We investigated the thermal phase structure and kinematics of atomic gas in
the Draco intermediate velocity cloud using DHIGLS H\,\textsc{i} 21~cm data
and ROHSA decomposition.
Our main results are summarized as follows.

\begin{enumerate}

\item
Two kinematically distinct WNM components are identified toward the Draco
region.
WNM1 is centered near $-3~\mathrm{km~s^{-1}}$ and is interpreted as local
ambient Galactic H\,\textsc{i}, whereas WNM2 is concentrated around
$-30~\mathrm{km~s^{-1}}$ and is associated with the infalling Draco IVC.
The LNM forms a broad velocity bridge between WNM1 and WNM2, and the CNM and
CO emission are confined to part of this intermediate velocity range.
This nested kinematic structure is consistent with the presence of a
dynamically compressed multiphase interaction layer produced by the
interaction between the infalling Draco cloud and ambient Galactic atomic gas.

\item
The H\,\textsc{i} mass fractions are $\sim41\%$, $\sim23\%$, $\sim17\%$,
and $\sim19\%$ for WNM1, WNM2, LNM, and CNM, respectively.
Although the WNM remains the dominant atomic phase, the LNM and CNM together
account for $\sim36\%$ of the total H\,\textsc{i} mass.
No comparably independent WNM component is recovered at intermediate
velocities, although a weak broad WNM component may remain unresolved.

\item
The observed phase structures are consistent with efficient thermal
processing of dynamically compressed WNM.
Compression increases the gas density and shortens the radiative cooling
time, allowing gas to evolve from the WNM through the thermally unstable LNM
toward the CNM.
In this interpretation, the LNM predominantly traces compressed and cooling
atomic gas, whereas the CNM represents regions in which cooling and
condensation have progressed further.

\item
We propose an evolutionary sequence of
``WNM $\rightarrow$ compressed atomic gas $\rightarrow$ LNM
$\rightarrow$ CNM $\rightarrow$ molecular gas.''
The present interaction may both create new CNM from WNM and further compress
or grow pre-existing dense structures.
These results provide the missing link between previous evidence for
dynamical interaction and molecular-cloud formation in Draco.
These results also suggest that CNM formation is a key step in transforming
gas accreting toward the Galactic disk into cold atomic and molecular
material.

\end{enumerate}

\section*{Funding}

This work was supported by JSPS KAKENHI Grant Numbers JP26K07137, JP25K23399, JP23H00129, and JP26H02068.

\section*{Acknowledgements}

We acknowledge the use of H\,\textsc{i} 21-cm data from the
DRAO H\,\textsc{i} Intermediate Galactic Latitude Survey
(DHIGLS; \citet{Blagrave2017}), which combines observations obtained
with the Dominion Radio Astrophysical Observatory Synthesis Telescope
and the Green Bank Telescope.
We also make use of the $^{12}\mathrm{CO}(J = 2\mbox{--}1)$ data of the
Draco nebula presented by \citet{BiegingKong2024}.
The UArizona ARO Submillimeter Telescope on Mt.\ Graham is operated by
the Arizona Radio Observatory (ARO), Steward Observatory,
University of Arizona.

\appendix

\section{Phase separation at different velocity resolutions}
\label{sec:velocity_resolution}

\begin{figure}
 \begin{center}
 \includegraphics[width=8cm]{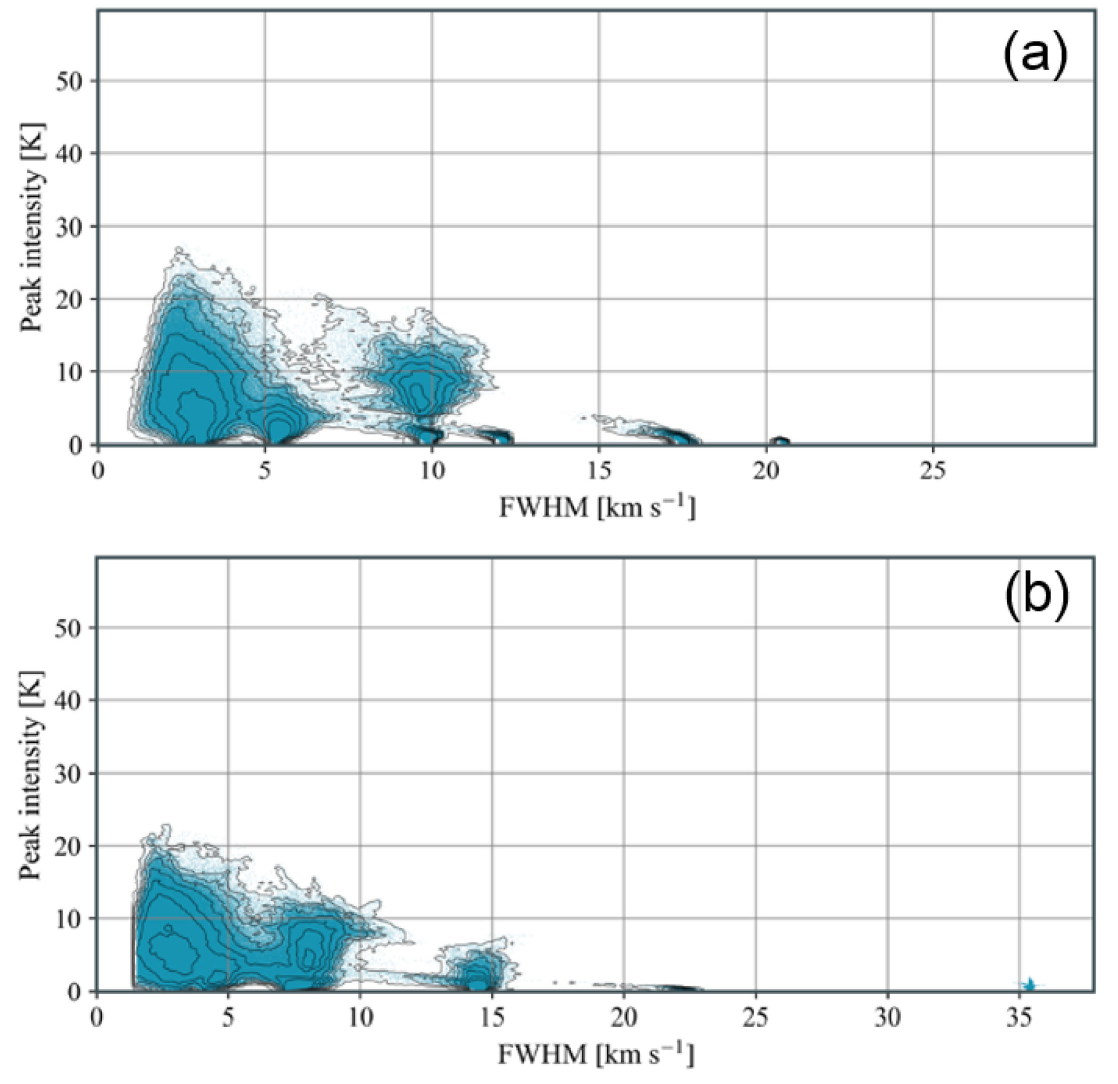}
 \end{center}
\caption{
Results of applying ROHSA to the GALFA-H\,\textsc{i} data of the
HLCG 92--35 region \citep{Matsuzuki2026}.
The H\,\textsc{i} data were smoothed to velocity resolutions of
(a) $\sim 0.36~\mathrm{km~s^{-1}}$ and
(b) $\sim 0.92~\mathrm{km~s^{-1}}$.\\
}
\label{fig:velocity_resolution}
\end{figure}

Figures~\ref{fig:velocity_resolution}(a) and (b) show the distributions of
the FWHMs and peak brightness temperatures of the individual Gaussian
components obtained by applying ROHSA to the GALFA-H\,\textsc{i} data used
by \citet{Matsuzuki2026}.
In their study, the analysis was performed using the original data with a
velocity resolution of $\sim 0.184~{\rm km~s^{-1}}$, in which the
distributions corresponding to the three H\,\textsc{i} phases are more
clearly distinguishable than those in Figure~\ref{fig:velocity_resolution}(a).

When the velocity resolution is degraded to
$\sim 0.9~{\rm km~s^{-1}}$, the components with the narrowest and
intermediate FWHMs begin to overlap.
This occurs because the degraded velocity resolution causes some
intrinsically narrow components, which can be properly resolved at higher
velocity resolution, to be measured with artificially broadened line widths.
Consequently, part of the CNM population overlaps with the LNM population
in the FWHM--peak brightness temperature plane.

These results indicate that a velocity resolution of
$\lesssim 0.4~{\rm km~s^{-1}}$ is required to reliably separate the
H\,\textsc{i} emission into the three thermal phases.

%%%%%%%%%%%%%%%%%%%%%%%%%%%%%%%%%%%%%%%

% Any journal's BST file (e.g., apj.bst) can be used as PASJ's BST is unavailable.    
%
% \bibliographystyle{****}
% \bibliography{****}
%

\end{document}